\documentclass[english,prl,reprint, superscriptaddress, showkeys, showpacs, raggedbottom, nobalancelastpage]{revtex4-1}
\usepackage[T1]{fontenc}
\usepackage[latin9]{inputenc}
\usepackage{xcolor}
\usepackage{pdfcolmk}
\usepackage{float}
\usepackage{units}
\usepackage{amsmath}
\usepackage{amssymb}
\usepackage{graphicx}
\usepackage{wasysym}
\PassOptionsToPackage{normalem}{ulem}
\usepackage{ulem}

\makeatletter

\providecolor{lyxadded}{rgb}{0,0,1}
\providecolor{lyxdeleted}{rgb}{1,0,0}
\usepackage{natbib,graphicx,url,times,layouts,xcolor,amsmath,amssymb,wasysym}
\usepackage{txfonts}   % For PRL fonts
\usepackage{cases}     % For numbering in cases format
\definecolor{blueviolet}{rgb}{0.2, 0.2, 0.6}
\usepackage[pdftex,          % FOR PDFLATEX ONLY
    bookmarks=false,          % Bookmark bar
    colorlinks=true,
    allcolors=blueviolet,
    pdfstartview={FitH},  % FitBH
    ]{hyperref}
\usepackage{pgffor}  % Including PDF
\usepackage{pdfpages}  % Including PDF

\makeatother

\usepackage{babel}
\begin{document}
\global\long\def\bra{\langle}
\global\long\def\ket{\rangle}
\global\long\def\half{\frac{1}{2}}
\global\long\def\p{\partial}
\global\long\def\a{\alpha}
\global\long\def\b{\beta}
\global\long\def\g{\gamma}
\global\long\def\c{\chi}
\global\long\def\d{\delta}
\global\long\def\o{\omega}
\global\long\def\m{\mu}
\global\long\def\n{\nu}
\global\long\def\z{\zeta}
\global\long\def\l{\lambda}
\global\long\def\e{\epsilon}
\global\long\def\x{\chi}
\global\long\def\r{\rho}
\global\long\def\t{\theta}
\global\long\def\G{\Gamma}
\global\long\def\D{\Delta}
\global\long\def\O{\Omega}
\global\long\def\L{\Lambda}
\global\long\def\P{\Phi}
\global\long\def\T{\Theta}
\global\long\def\dg{\dagger}
\global\long\def\s{\sigma}
\global\long\def\dag{\dagger}
\global\long\def\ma{\left|\a\right|}
\global\long\def\A{\mathcal{A}}
\global\long\def\ct#1{|#1_{\a}\rangle}
\global\long\def\aa{\hat{a}}
\global\long\def\bb{\hat{b}}
\global\long\def\cc{\hat{c}}
\global\long\def\rr{\hat{\rho}}
\global\long\def\pr{\hat{\pi}}
\global\long\def\ph{\hat{n}}
\global\long\def\ff{F}
\global\long\def\rot{R}
\global\long\def\disp{D}
\global\long\def\proj{\Pi}
\global\long\def\ssproj{P}
\global\long\def\pos{\hat{\chi}}
\global\long\def\gen{\hat{\pi}_{\n}}
\global\long\def\elle{\lambda}
\global\long\def\SS{S}
\global\long\def\RR{R}
\global\long\def\bb{\hat{b}}
\global\long\def\k{\kappa}
\global\long\def\lket#1{\left|#1\right\rangle }
\DeclareRobustCommand{\bra}{\langle}
\DeclareRobustCommand{\ket}{\rangle}
\renewcommand{\sec}[1]{\textit{#1.---}}
\renewcommand{\rmdefault}{ptm}
\newcommand{\hqcdfs}{carollo2006a,*zhang2006,*Yin2007,*dasgupta2007,*feng2009,*zheng2012,*zheng2014}
\newcommand{\hqcns}{oreshkov2009,*oreshkov2009a,oreshkov2010}
\newcommand{\nadfs}{xu2012,*wu2012,*xu2014,*liang2014,*mousolou2014,*pyshkin2015,*Xue2016}
\newcommand{\nans}{zhang2014a}%\renewenvironment{lyxgreyedout}{\textcolor{blue}\bgroup}{\egroup}

\title{Holonomic quantum control with continuous variable systems}

\author{Victor~V.~Albert}

\affiliation{Departments of Applied Physics and Physics, Yale University, New
Haven, Connecticut, USA}

\author{Chi~Shu}

\affiliation{Departments of Applied Physics and Physics, Yale University, New
Haven, Connecticut, USA}

\affiliation{Department of Physics, The Hong Kong University of Science and Technology,
Hong Kong, China}

\author{Stefan~Krastanov}

\affiliation{Departments of Applied Physics and Physics, Yale University, New
Haven, Connecticut, USA}

\author{Chao~Shen}

\affiliation{Departments of Applied Physics and Physics, Yale University, New
Haven, Connecticut, USA}

\author{Ren-Bao~Liu}

\affiliation{Department of Physics \& Centre for Quantum Coherence, The Chinese
University of Hong Kong, Hong Kong, China}

\author{Zhen-Biao~Yang}

\affiliation{Departments of Applied Physics and Physics, Yale University, New
Haven, Connecticut, USA}

\affiliation{Department of Physics, Fuzhou University, Fuzhou, China}

\author{Robert~J.~Schoelkopf}

\affiliation{Departments of Applied Physics and Physics, Yale University, New
Haven, Connecticut, USA}

\author{Mazyar~Mirrahimi}

\affiliation{Departments of Applied Physics and Physics, Yale University, New
Haven, Connecticut, USA}

\affiliation{INRIA Paris-Rocquencourt, Domaine de Voluceau, Le Chesnay Cedex,
France}

\author{Michel~H.~Devoret}

\affiliation{Departments of Applied Physics and Physics, Yale University, New
Haven, Connecticut, USA}

\author{Liang~Jiang}

\affiliation{Departments of Applied Physics and Physics, Yale University, New
Haven, Connecticut, USA}

\pacs{03.65.Yz, 03.65.Vf, 42.50.Dv}

\keywords{Lindblad master equation, decoherence free subspace, holonomy, reservoir
engineering}

\date{\today}
\begin{abstract}
Universal computation of a quantum system consisting of superpositions
of well-separated coherent states of multiple harmonic oscillators
can be achieved by three families of adiabatic holonomic gates. The
first gate consists of moving a coherent state around a closed path
in phase space, resulting in a relative Berry phase between that state
and the other states. The second gate consists of ``colliding''
two coherent states of the same oscillator, resulting in coherent
population transfer between them. The third gate is an effective controlled-phase
gate on coherent states of two different oscillators. Such gates should
be realizable via reservoir engineering of systems which support tunable
nonlinearities, such as trapped ions and circuit QED.
\end{abstract}
\maketitle
Reservoir engineering schemes continue to reveal promising new directions
in the search for potentially robust and readily realizable quantum
memory platforms. Such schemes are often described by \textit{Lindbladians}
\cite{gorini1976a,*lindblad1976} possessing \textit{decoherence-free
subspaces} (DFSs) \cite{duan1997,*Zanardi1997,*lidar1998} or (more
generally) \textit{noiseless subsystems} (NSs) \cite{knill2000, *zanardi2000, *kempe2001}
-- multidimensional spaces immune to the nonunitary effects of the
Lindbladian and, potentially, to other error channels \cite{zanardi2015,cats}.
On the other hand, \textit{holonomic quantum computation} (HQC) \cite{zanardi1999,*pachos1999,*lidarbook_zanardi}
is a promising framework for achieving noise-resistant quantum computation
\cite{Solinas2004}. In HQC, states undergo adiabatic closed-loop
parallel transport in parameter space, acquiring Berry phases or matrices
(also called non-Abelian holonomies or Wilson loops \cite{wilczek1984})
which can be combined to achieve universal computation.

\begin{figure}[!t]
\centering{}\includegraphics[width=1\columnwidth]{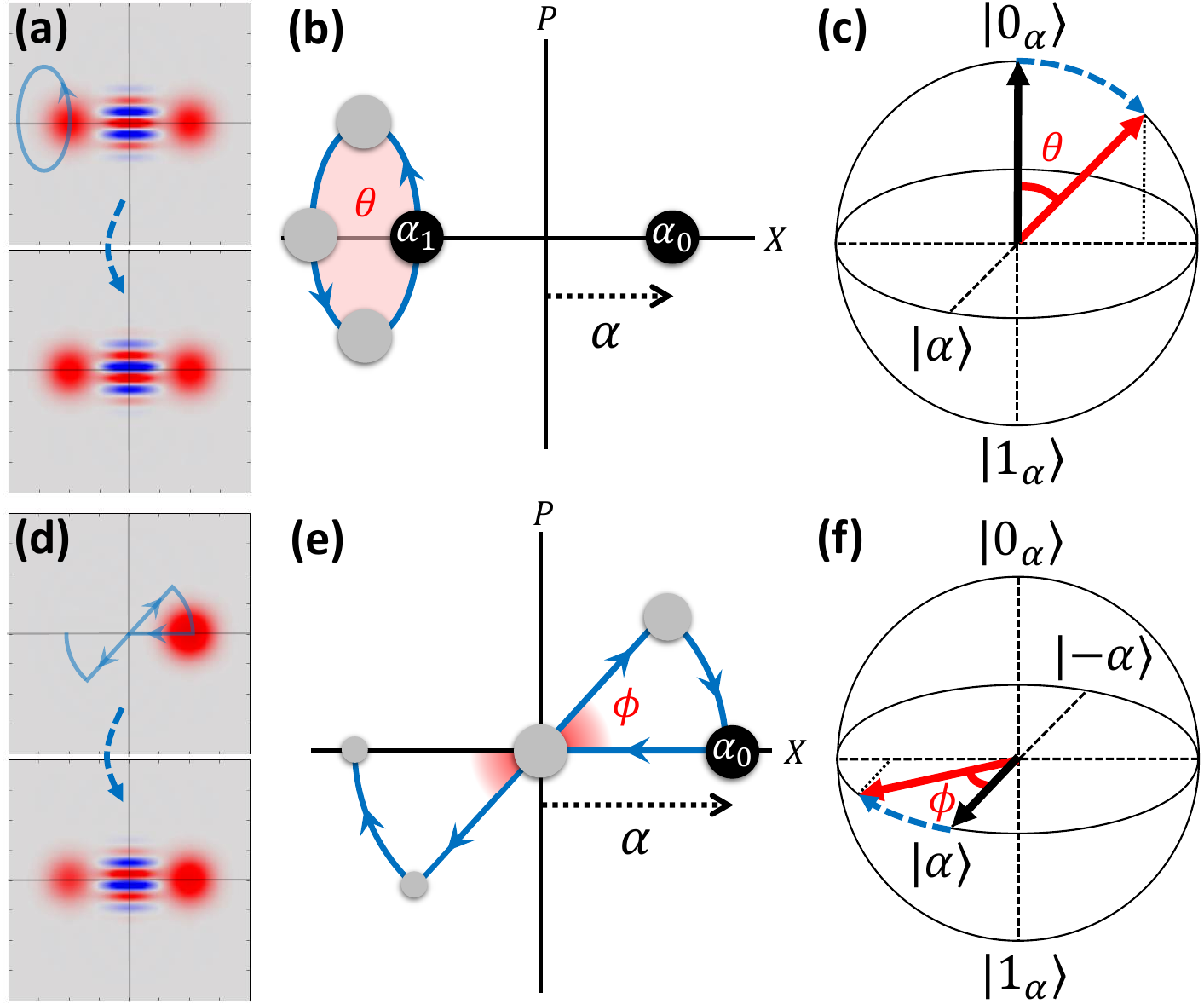}\caption{\label{f1}In the $d=2$ cat-code, quantum information is encoded
in the coherent states $|\protect\a_{0}(0)\protect\ket\equiv|\protect\a\protect\ket$
and $|\protect\a_{1}(0)\protect\ket\equiv\protect\lket{-\protect\a}$.
\textbf{(a)} Wigner function sketch of the state before (top) and
after (bottom) a loop gate acting on $\protect\lket{-\protect\a}$,
depicting the path of $\protect\lket{-\protect\a}$ during the gate
(blue) and a shift in the fringes between $\protect\lket{\pm\protect\a}$.
\textbf{(b)} Phase space diagram for the loop gate; $X=\frac{1}{2}\protect\bra\protect\aa+\protect\aa^{\protect\dg}\protect\ket$
and $P=-\frac{i}{2}\protect\bra\protect\aa-\protect\aa^{\protect\dg}\protect\ket$.
The parameter $\protect\a_{1}(t)$ is varied along a closed path (blue)
of area $A$, after which the state $\protect\lket{-\protect\a}$
gains a phase $\protect\t=2A$ relative to $|\protect\a\protect\ket$.
\textbf{(c)} Effective Bloch sphere of the $\protect\lket{\pm\protect\a}$
qubit depicting the rotation caused by the $d=2$ loop gate. Black
arrow depicts initial state while red arrow is the state after application
of the gate. The dotted blue arrow does not represent the path traveled
since the states leave the logical space $\protect\lket{\pm\protect\a}$
during the gate.\textbf{ (d-f)} Analogous descriptions of the collision
gate, which consists of reducing $\protect\a$ to 0, driving back
to $\protect\a\exp(i\phi)$, and rotating back to $\protect\a$.}
\end{figure}

It is natural to consider combining the above two concepts. After
the initial proposals \cite{zanardi2001,pachos2004}, the idea of
HQC on a DFS gained traction in Refs. \cite{wu2005,carollo2006} and
numerous investigations into HQC on DFSs \cite{\hqcdfs} and NSs \cite{\hqcns,abfj}
followed. However, previous proposals perform HQC on DFS states constructed
out of a finite-dimensional basis of atomic or \textit{spin} states.
There has been little investigation \cite{Lucarelli2002} of HQC on
DFSs consisting of nontrivial oscillator states (e.g. coherent states
\cite{Cacheffo2010,*Neto2011,DePonte2007, *DePonte2008, *DePonte2011}).
While this is likely due to a historically higher degree of control
of spin systems, recent experimental progress in control of microwave
cavities \cite{sayrin2011, *Kirchmair2013, *Vlastakis2013, *sun2014, *heeres2015, leghtas2014,ofek2016},
trapped ions \cite{zoller_stabilizers, *Schindler2013}, and Rydberg
atoms \cite{Signoles2014} suggests that oscillator-type systems are
also within reach. In this Letter, we propose an oscillator HQC-on-DFS
scheme using cat-codes.

Cat-codes are quantum memories for coherent-state quantum information
processing \cite{catbook,*cvbook_cats} storing information in superpositions
of well-separated coherent states which are evenly distributed around
the origin of phase space. Cat-code quantum information can be protected
from cavity dephasing via passive quantum error correction \cite{terhal2015}
using Lindbladian-based reservoir engineering \cite{cats}. In addition,
such information can be actively protected from photon loss events
\cite{Cochrane1999,Glancy2004,Leghtas2013b,cats}. While there exist
plenty of methods to create and manipulate the necessary states \cite{jeong2002,*Ralph2003,*gilchrist2004,*Ourjoumtsev2006,*Takeoka2007,*lund2008,*Takahashi2008,Cochrane1999,Munro2000,arenz2013,*coolforcats,cats}
and while the gates can also be implemented using Hamiltonians, we
consider reservoir engineering due to its protective features. Cat-codes
differ from the well-known Gottesman-Kitaev-Preskill (GKP) encoding
scheme \cite{Gottesman2001} in both state structure and protection.
GKP codes consist of superpositions of highly squeezed states and
focus on protecting against small shifts in oscillator position and
momentum. In contrast, cat codes protect against damping and dephasing
errors, the dominant loss mechanisms for most cavity systems. While
realistic GKP realization schemes remain scarce \cite{terhal2015b},
cat-codes benefit from greater near-term experimental feasibility
\cite{ofek2016}.

For simplicity, let us introduce our framework using a single oscillator
(or mode). Consider the Lindbladian
\begin{equation}
\dot{\rho}=\ff\r\ff^{\dg}-\half\{\ff^{\dg}\ff,\r\}\,\,\,\,\,\,\,\,\,\text{ with }\,\,\,\,\,\,\,\,\,\ff=\sqrt{\k}\prod_{\n=0}^{d-1}(\aa-\a_{\n})\,,\label{eq:main}
\end{equation}
$[\aa,\aa^{\dg}]=1,$ $\ph\equiv\aa^{\dg}\aa$, $\k\in\mathbb{R}$,
dimensionless $\a_{\n}\in\mathbb{C}$, and $\r$ a density matrix.
The $d=1$ case {[}$\ff=\sqrt{\k}(\aa-\a_{0})${]} reduces to the
well-known driven damped harmonic oscillator (\cite{klimov_book},
Sec. 9.1) whose unique steady state is the coherent state $|\a_{0}\ket$
(with $\aa|\a_{0}\ket=\a_{0}|\a_{0}\ket$). Variants of the $d=2$
case are manifest in driven 2-photon absorption (\cite{puri}, Sec.
13.2.2), the degenerate parametric oscillator (\cite{carmichael2},
Eq.~12.10), and a laser-driven trapped ion (\cite{poyatos1996},
Fig.~2d; see also \cite{Garraway1998,*carvalho2001}). A motivation
for this work has been the recent realization of the $\ff=\sqrt{\k}(\aa^{2}-\a_{0}^{2})$
process in circuit QED \cite{leghtas2014}, following an earlier proposal
to realize $\ff=\sqrt{\k}(\aa^{d}-\a_{0}^{d})$ with $d=2,4$ \cite{cats}.
For arbitrary $d$ and certain $\a_{\n}$, a qu$d$it steady state
space is spanned by the $d$ well-separated coherent states $|\a_{\n}\ket$
that are annihilated by $\ff$. The main conclusion of this work is
that universal control of this qudit can be done via two simple gate
families, \textit{loop gates} and \textit{collision gates}, that rely
on adiabatic variation of the parameters $\a_{\n}(t)$. Universal
computation on multiple modes can then be achieved with the help of
an entangling two-oscillator \textit{infinity gate}. We first sketch
the $d=2$ case and extend to arbitrary $d$ with $|\a_{\n}\ket$
arranged in a circle in phase space. The straightforward generalization
to arbitrary arrangements of $|\a_{\n}\ket$ is presented in \footnote{See Supplemental Material, which cites \cite{davies1978,*Thunstrom2005,*pekola2010,Lloyd1995,Avron2012a,Sarandy2005,sarandy2006,Avron2011,*Avron2012b,Venuti2015,\nadfs,\nans,braunstein2005,beige2000a,*burgarth2013,facchi2002,*paz-silva2012},
at URL for an extension of the infinity gate to qudits, proof of universal
computation, generalization to arbitrary $|\a_{\n}\ket$, a sketch
of the derivation of the Berry matrices, and details on how to integrate
the gates with a photon loss error correction scheme.}. We then discuss gate errors and integration with cat-code error
correction schemes \cite{Leghtas2013b,cats}, concluding with a discussion
of experimental implementation.

\sec{Single-qubit gates}Let $d=2$ and let $\a_{0},\a_{1}$ depend
on time in Eq.~(\ref{eq:main}), so the steady-state space holds
a qubit worth of information. The positions of the qubit's two states
$|\a_{\n}(t)\ket$ in phase space are each controlled by a tunable
parameter. We let $\a_{0}(0)=-\a_{1}(0)\equiv\a$ (with $\a$ real
unless stated otherwise). This system's steady states $\lket{\pm\a}$
are the starting point of parameter space evolution for this section
and the qubit defined by them (for large enough $\a$) is shown in
Fig.~\ref{f1}a.

The loop gate involves an adiabatic variation of $\a_{1}(t)$ through
a closed path in phase space (see Fig.~\ref{f1}b). The state $|\a_{1}(t)\ket$
will follow the path and, as long as the path is well separated from
$|\a_{0}(t)\ket=|\a\ket$, will pick up a phase $\t=2A$, with $A$
being the area enclosed by the path \cite{Chaturvedi1987}. It should
be clear that initializing the qubit in $\lket{-\a}$ will produce
only an irrelevant \textit{overall} phase upon application of the
gate (similar to the $d=1$ case). However, once the qubit is initialized
in a superposition of the two coherent states with coefficients $c_{\pm}$,
the gate will impart a \textit{relative} phase: 
\begin{equation}
c_{+}|\a\ket+c_{-}\lket{-\a}\longrightarrow c_{+}|\a\ket+c_{-}e^{i\theta}\lket{-\a}\,.
\end{equation}
Hence, if we pick $|\a\ket$ to be the $x$-axis of the $\lket{\pm\a}$
qubit Bloch sphere, this gate can be thought of as a rotation around
that axis (depicted blue in Fig.~\ref{f1}c). Similarly, adiabatically
traversing a closed and isolated path with the other state parameter
$|\a_{0}(t)\ket$ will induce a phase on $|\a\ket$.

We now introduce the remaining Bloch sphere components of the cat-code
qubit. For $\a=0$, the $d=2$ case retains its qubit steady-state
space, which now consists of Fock states $|\m\ket$, $\m=0,1$ (since
$\ff=\sqrt{\k}\aa^{2}$ annihilates both). One may have noticed that
both states $\lket{\pm\a}$ go to $|0\ket$ in the $\a\rightarrow0$
limit and do not reproduce the $\a=0$ steady state basis. This issue
is resolved by introducing the cat state basis \cite{dodonov1974}
\begin{equation}
\ct{\m}\equiv\frac{e^{-\half\a^{2}}}{\mathcal{N}_{\m}}\sum_{n=0}^{\infty}\frac{\a^{2n+\m}}{\sqrt{(2n+\m)!}}|2n+\m\ket\overset{_{^{\a\rightarrow\infty}}}{\,\sim\,}\frac{1}{\sqrt{2}}(|\a\ket+(-)^{\m}\lket{-\a})\label{eq:twocat}
\end{equation}
with normalization $\mathcal{N}_{\m}=\sqrt{\half[1+(-)^{\m}\exp(-2\a^{2})]}$.
As $\a\rightarrow0$, $\ct{\m}\sim|\m\ket$ while for $\a\rightarrow\infty$,
the cat states (exponentially) quickly become ``macroscopic'' superpositions
of $\lket{\pm\a}$. This problem thus has \textit{only two distinct
parameter regimes}: one in which coherent states come together ($\a\ll1$)
and one in which they are well-separated ($\a\gg1,$ or more practically
$\a\apprge2$ for $d=2$). Eq.~(\ref{eq:twocat}) shows that (for
large enough $\a$) cat states and coherent states become conjugate
$z$- and $x$-bases respectively, forming a qubit. We note that $\m=0,1$
labels the respective $\pm1$ eigenspace of the parity operator $\exp(i\pi\ph)$;
this photon parity is preserved during the collision gate.

We utilize the $\a\ll1$ regime to perform rotations around the Bloch
sphere $z$-axis (Fig.~\ref{f1}f), which effectively induce a collision
and population transfer between $|\a\ket$ and $\lket{-\a}$. The
procedure hinges on the following observation: applying a bosonic
rotation $\rot_{\phi}\equiv\exp(i\phi\ph)$ to well-separated coherent
or cat state superpositions \textit{does not} induce state-dependent
phases while applying $\rot_{\phi}$ to Fock state superpositions
\textit{does}. Only one tunable parameter $\a_{0}(t)=-\a_{1}(t)$
is necessary here, so $\ff=\sqrt{\k}[\aa^{2}-\a_{0}(t)^{2}]$ with
$|\a_{0}(0)|=\a$. The collision gate consists of reducing $\a$ to
0, driving back to $\a\exp(i\phi)$, and rotating back to $\a$ (Fig.~\ref{f1}e).
The full gate is thus represented by $\RR_{\phi}^{\dg}\SS_{\phi}\SS_{0}^{\dg}$,
with $\SS_{\phi}$ \footnote{When acting on states in the DFS, $S_{\phi}$ can be approximated
by a path-ordered product of DFS projections $\ssproj_{\a}=|0_{\a}\ket\bra0_{\a}|+|1_{\a}\ket\bra1_{\a}|$
with each projection incrementing $\a$: $S_{\phi}\approx P_{\a e^{i\phi}}\cdots P_{\frac{2}{M}\a e^{i\phi}}P_{\frac{1}{M}\a e^{i\phi}}$
for integer $M\gg1$ \cite{zanardi2015}. Using Eq.~(\ref{eq:twocat}),
one can show that $\ssproj_{\a e^{i\phi}}=\rot_{\phi}\ssproj_{\a}\rot_{\phi}^{\dg}$
and prove Eq.~(\ref{eq:conj}).} denoting the nonunitary driving from 0 to $\a\exp(i\phi)$ . Since
\begin{equation}
\RR_{\phi}^{\dg}\SS_{\phi}\SS_{0}^{\dg}=\RR_{\phi}^{\dg}(\RR_{\phi}\SS_{0}\RR_{\phi}^{\dg})\SS_{0}^{\dg}=\SS_{0}\RR_{\phi}^{\dg}\SS_{0}^{\dg}\,,\label{eq:conj}
\end{equation}
the collision gate is equivalent to reducing $\a$, applying $\rot_{\phi}^{\dg}$
on the steady-state basis $|\m\ket$, and driving back to $\a$. The
net result is thus a relative phase between the states $\ct{\m}$:
\begin{equation}
c_{0}\ct 0+c_{1}\ct 1\longrightarrow c_{0}\ct 0+c_{1}e^{-i\phi}\ct 1\,.
\end{equation}
In the coherent state basis, this translates to a coherent population
transfer between $\lket{\pm\a}$.

\begin{figure}[H]
\centering{}\includegraphics[width=1\columnwidth]{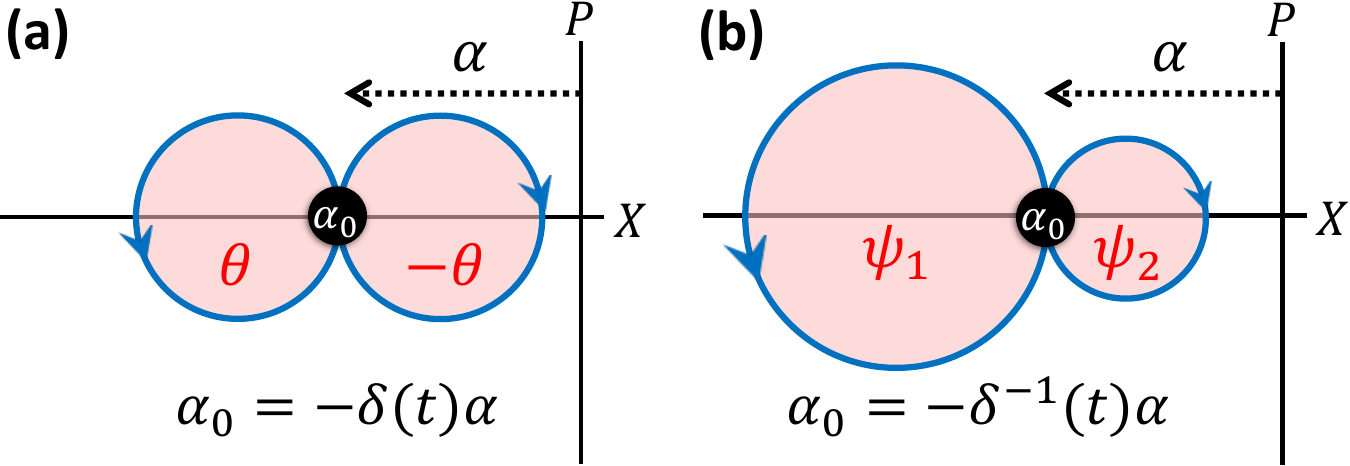}\caption{\label{f1b}Sketch of the adiabatic paths of the components \textbf{(a)}
$\protect\lket{-\protect\d\protect\a}$ and \textbf{(b)} $|\!-\!\protect\d^{-1}\protect\a\protect\ket$
during the infinity gate.}
\end{figure}

\sec{Two-qubit gates}Now let's add a second mode $\bb$ and introduce
the entangling infinity gate for the $2$-photon case. We now use
two jump operators for Eq. (\ref{eq:main}),
\begin{equation}
\ff_{I}=\left(\aa-\a\right)\left(\aa+\d\a\right)\,\,\,\,\,\,\text{and}\,\,\,\,\,\,\ff_{II}=(\aa\bb-\a^{2})(\aa\bb+\d\a^{2})\,.\label{eq:twomode}
\end{equation}
We keep $\a>0$ constant and vary $\d(t)$ in a figure-eight or ``$\infty$''
pattern (Fig. \ref{f1b}), starting and ending with $\d=1$. For $\d=1$,
the four DFS basis elements $\{\lket{\pm\a}\}\otimes\{\lket{\pm\a}\}$
are annihilated by both $\ff_{I}$ and $\ff_{II}$. For $\d\neq1$
and for sufficiently large $\a$, the basis elements become $\lket{\a,\a}$,
$\lket{\a,-\d\a}$, $\lket{-\d\a,\a}$, and $|-\d\a,-\d^{-1}\a\ket$.
Notice that the $\d^{-1}$ makes sure that $F_{II}|\!-\!\d\a,-\d^{-1}\a\ket=0$.
This $\d^{-1}$ allows the fourth state to gain a Berry phase distinct
from the other three states. Since Berry phases of different modes
add, we analyze the $\aa$/$\bb$-mode contributions individually.
For any state which contains the $\lket{-\d\a}$ component (in either
mode), the Berry phase gained for each of the two circles is proportional
to their areas. Since the oppositely oriented circles have the same
area (Fig. \ref{f1b}a), these phases will cancel. The Berry phase
of the fourth state, which contains the component $|\!-\!\d^{-1}\a\ket$,
will be proportional to the total area enclosed by the path made by
$\d^{-1}$. Inversion maps circles to circles, but the two inverted
circles will now have \textit{different} areas (Fig. \ref{f1b}b).
Summing the Berry phases $\psi_{i}$ gained upon traversal of the
two circles $i\in\{1,2\}$ yields an effective phase gate:
\begin{equation}
\lket{-\a,-\a}+|\text{rest}\ket\longrightarrow e^{i(\psi_{1}+\psi_{2})}\lket{-\a,-\a}+|\text{rest}\ket\,,
\end{equation}
where $|\text{rest}\ket$ is the unaffected superposition of the remaining
components $\{\lket{\a,\a},\lket{\a,-\a},\lket{-\a,\a}\}$.

\begin{figure}[h]
\centering{}\includegraphics[width=1\columnwidth]{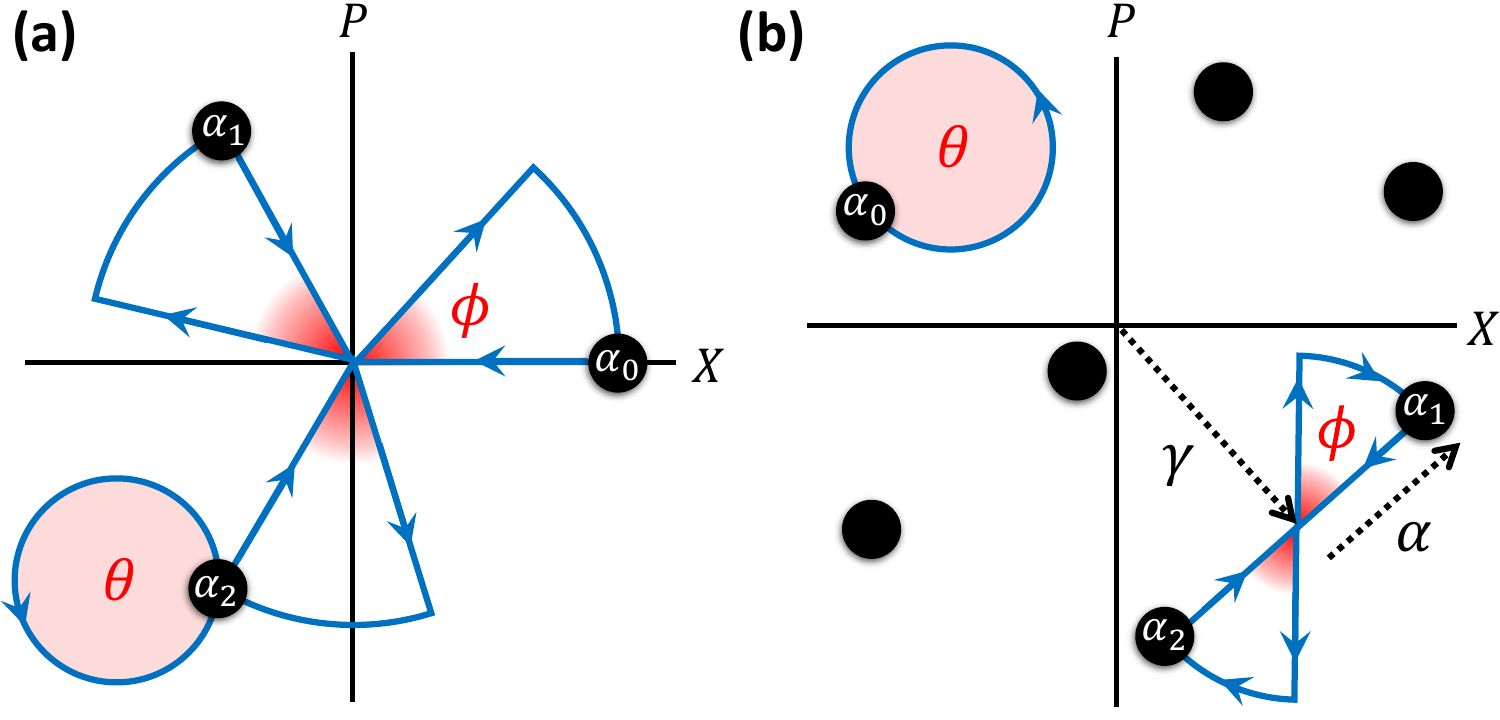}\caption{\label{f2}\textbf{(a)} Three-fold symmetric configuration of steady
states $|\protect\a_{\protect\n}\protect\ket$ of Eq.~(\ref{eq:main})
with $d=3$ and depiction of a loop gate ($\theta$) acting on $|\protect\a_{2}\protect\ket$
and a collision gate $(\phi$) between all states. \textbf{(b)} Arbitrary
configuration of steady states for $d=7$, depicting $|\protect\a_{0}\protect\ket$
undergoing a loop gate and $|\protect\a_{1}\protect\ket,|\protect\a_{2}\protect\ket$
undergoing a displaced collision gate (see Supplementary Material).}
\end{figure}

\sec{Single-qudit gates}We now outline the system and its single-mode
gates for arbitrary $d$. Here we let $\a_{\n}(0)\equiv\a e_{\n}$
with real non-negative $\a$, $e_{\n}\equiv\exp(i\frac{2\pi}{d}\n)$,
and $\n=0,1,\cdots,d-1$ (see Fig.~\ref{f2}a for $d=3$). This choice
of initial qudit configuration makes Eq.~(\ref{f1}) invariant under
the discrete rotation $\exp(i\frac{2\pi}{d}\ph)$ and is a bosonic
analogue of a particle on a discrete ring \cite{vourdas2004}. Therefore,
$\ph\text{mod}d$ is a good quantum number and we can distinguish
eigenspaces of $\exp(i\frac{2\pi}{d}\ph)$ by projections \cite{pub011}
\begin{equation}
\proj_{\m}=\sum_{n=0}^{\infty}|dn+\m\ket\bra dn+\m|=\frac{1}{d}\sum_{\n=0}^{d-1}\exp\left[i\frac{2\pi}{d}(\ph-\m)\n\right]
\end{equation}
with $\m=0,1,\cdots,d-1$. The corresponding cat-state basis generalizes
Eq.~(\ref{eq:twocat}) to\begin{subnumcases}{\!\!\!\!\!\!\!\!\!\ct{\m}\equiv\frac{\proj_{\m}|\a\ket}{\sqrt{\bra\a|\proj_{\m}|\a\ket}}\sim\!}
\!\!\! |\m\ket & $\! \a\rightarrow0$ \label{eq:gencats1}\\   
\!\!\! \textstyle{\frac{1}{\sqrt{d}}\sum_{\n=0}^{d-1}e^{-i\frac{2\pi}{d}\m\n}|\a e_{\n}\ket} & $\! \a\rightarrow\infty$ .\label{eq:gencats2}
\end{subnumcases}Since overlap between coherent states decays exponentially with $\a$,
the quantum Fourier transform between coherent states $|\a e_{\n}\ket$
and cat states $\ct{\m}$ in Eq.~(\ref{eq:gencats2}) is valid in
the well-separated regime, i.e., when $2\a\sin\frac{\pi}{d}\gg1$
(satisfied when $|\bra\a|\a e_{1}\ket|^{2}\ll1$). It should be clear
that the more coherent states there are (larger $d$), the more one
has to drive to resolve them (larger $\a$). Also note the proper
convergence to Fock states $|\m\ket$ as $\a\rightarrow0$ in Eq.~(\ref{eq:gencats1}).

Both gates generalize straightforwardly (see Fig.~\ref{f2}a for
$d=3$). The loop gate consists of adiabatic evolution of a specific
$\a_{\n}(t)$ around a closed path isolated from all other $\a_{\n^{\prime}}(0)$.
There are $d$ such possible evolutions, each imparting a phase on
its respective $|\a e_{\n}\ket$. The collision gate is performed
as follows: starting with the $|\a e_{\n}\ket$ configuration for
large enough $\a$, tune $\a$ to zero (or close to zero), pump back
to a different phase $\a\exp(i\phi)$, and rotate back to the initial
configuration. Each $|\m_{\a}\ket$ will gain a phase proportional
to its mean photon number, which behaves in the two parameter regimes
as follows:\begin{subnumcases}{\bra\m_{\a}|\ph|\m_{\a}\ket = }
\m+O(\a^{2d}) & $\a\rightarrow0 \label{eq:gauge1}$ \\
\a^{2}+O(\a^{2}e^{-c\a^{2}}) & $\a\rightarrow\infty$ , \label{eq:gauge2} 
\end{subnumcases}where $c=1-\cos\nicefrac{2\pi}{d}$. Since a rotation imparts only
a $\m$-independent (i.e. overall) phase in the well-separated regime
{[}Eq.~(\ref{eq:gauge2}){]}, the only $\m$-dependent (i.e. nontrivial)
contribution of the symmetric collision gate path to the Berry matrix
is at $\a=0$. This gate therefore effectively applies the Berry matrix
$\exp(-i\phi\pos)$ to the qudit, where $\pos\equiv\sum_{\m=0}^{d-1}\m\ct{\m}\bra\m_{\a}|$
is the discrete position operator of a particle on a discrete ring
\cite{vourdas2004}. More generally, one does not have to tune $\a$
all the way to zero to achieve similar gates -- e.g. being in the
regime with $2\a\sin\frac{\pi}{d}\approx1$ is sufficient. The two-mode
infinity gate can likewise be extended to the $d$-photon case and
it is a simple exercise to prove universality \protect\footnotemark[1].

\sec{Gate errors} In the Lindbladian-dominated adiabatic limit \footnotemark[1],
the role of the excitation gap is played by the \textit{dissipation
gap} -- the eigenvalue of the Lindblad operator whose real part is
closest (but not equal) to zero. Since our Lindbladians are infinite-dimensional,
it is possible for the dissipation gap to approach zero for sufficiently
large $|\a_{\n}|$ (i.e., in the limit of an infinite-dimensional
space). For the symmetric $d-$photon case however, this is not the
case and the gap actually \textit{increases} with $\a$ (verified
numerically for $d\leq10$). Having numerically verified the infinity-gate,
we also see that the gap increases with $\a$ in the two-mode system
(\ref{eq:twomode}). The gap can also be seen to increase by analyzing
the excitation gap of the Hamiltonian $F^{\dg}F$ (see Ref. \cite{abfj},
Sec. VIII.C).

Here we discuss the scaling of leading-order nonadiabatic errors,
focusing on the single-mode gates for $d=2,3$.

Non-adiabatic corrections in Lindbladians are in general nonunitary,
so their effect is manifest in the \textit{impurity} of the final
state (assuming a pure initial state). Extensive numerical simulations
\cite{ani} show that the impurity can be fit to
\begin{equation}
\epsilon\equiv1-\text{Tr}\{\r(T)^{2}\}\propto\frac{1}{\k T\a^{p}}
\end{equation}
as $\a,T\rightarrow\infty$, where $\r(T)$ is the state after completion
of the gate, $\a\equiv|\a_{\n}(0)|$ is the initial distance of all
$|\a_{\n}\ket$ from the origin, $p>0$ is gate-dependent, and $\k$
is the overall rate of Eq. (\ref{eq:main}). One can see that $\epsilon\approx O(T^{-1})$,
as expected for a nearly adiabatic process. Additionally, we report
that $p\approx1.8$ for $d=2$ and $p\approx3.9$ for $d=3$ loop
gates, respectively. For the $d=2,3$ collision gates, we observe
that $p\approx0$. 

\sec{Photon loss errors}We have determined that the above gates can
be made compatible with a (photon number) parity-based scheme protecting
against photon loss \cite{Leghtas2013b,cats}. In such a scheme, one
encodes quantum information in a logical space spanned by even parity
states (e.g. $\lket{\a_{\n}}+\lket{-\a_{\n}}$ with $\nu=0,1,\cdots,d-1$,
generalizing Sec. II.D.3 of \cite{terhal2015}). Photon loss events
can be detected by quantum non-demolition measurements of the parity
operator $(-1)^{\ph}$. In the case of fixed-parity cat-codes, errors
due to photon loss events can be corrected immediately \cite{Leghtas2013b}
or tracked in parallel with the computation \cite{cats}. By doubling
the size $d$ of the DFS of Lindbladian (\ref{eq:main}) to accomodate
both even and odd parity logical spaces, we have determined a set
of holonomic gates which are parity conserving and are universal on
each parity subspace \protect\footnotemark[1]. This allows for parity
detection to be performed before/after HQC.

\sec{Implementation \& conclusion}We show how to achieve universal
computation of an arbitrary configuration of multi-mode well-separated
coherent states $|\a_{\n}\ket$ by adiabatic closed-loop variation
of $\a_{\n}(t)$. We construct Lindbladians which admit a decoherence-free
subspace consisting of such states and whose jump operators consist
of lowering operators of the modes. One can obtain the desired jump
operators by nonlinearly coupling the multi-mode system to auxiliary
modes ($\cc,\hat{d},\cdots$), which act as effective thermal reservoirs
for the active modes. For the case of one active mode $\aa$, if one
assumes a coupling of the form $\aa\cc^{\dg}+H.c.$ and no thermal
fluctuations in $\cc$, one will obtain (in the Born-Markov approximation)
a Lindbladian with jump operator $\aa$. Therefore, a generalization
of the coupling to $F\cc^{\dg}+H.c.$ will result in the desired single-mode
Lindbladian (\ref{eq:main}) with jump operator $F$. Since $F$ are
polynomials in the lowering operators of the active modes, quartic
and higher mode interactions need to be engineered. Such terms can
be obtained by driving an atom in a harmonic trap with multiple lasers
\cite{poyatos1996} or by coupling between a Josephson junction and
a microwave cavity \cite{cats, leghtas2014}. We thus describe arguably
the first approach to achieve holonomic quantum control of realistic
continuous variable systems. 
\begin{acknowledgments}
The authors thank S. M. Girvin, L. I. Glazman, N. Read, and Z. Leghtas
for fruitful discussions. This work was supported by the ARO, AFOSR
MURI, DARPA Quiness program, the Alfred P. Sloan Foundation, and the
David and Lucile Packard Foundation. V.V.A. was supported by the NSF
GRFP under Grant DGE-1122492. C.S. was supported by the Paul and May
Chu Overseas Summer Research Travel Grant from HKUST and acknowledges
support from Shengwang Du. R.-B. L. was supported by Hong Kong RGC/GRF
Project 14303014. We thank the Yale High Performance Computing Center
for use of their resources.
\end{acknowledgments}

\bibliographystyle{apsrev4-1}
\bibliography{C:/Users/Victor/library}
%\includepdf[]{cats_supp.pdf}


\section*{Supplementary material}

# Supplemental material: Universal holonomic quantum computing with cat-codes

Victor V. Albert,[1] Chi Shu,[1,2] Stefan Krastanov,[1] Chao Shen,[1] Ren-Bao Liu,[3] Zhen-Biao Yang,[1,4]
Robert J. Schoelkopf,[1] Mazyar Mirrahimi,[1,5] Michel H. Devoret,[1] and Liang Jiang[1]

[1]*Departments of Applied Physics and Physics, Yale University, New Haven, Connecticut, USA*
[2]*Department of Physics, The Hong Kong University of Science and Technology, Hong Kong, China*
[3]*Department of Physics & Centre for Quantum Coherence,*
*The Chinese University of Hong Kong, Hong Kong, China*
[4]*Department of Physics, Fuzhou University, Fuzhou, China*
[5]*INRIA Paris-Rocquencourt, Domaine de Voluceau, Le Chesnay Cedex, France*
(Dated: March 30, 2016)

**Summary**

In Sec. A, we provide a qudit extension of the two-qubit infinity gate described in the main text. In Sec. B, we prove universal quantum computation on the subspace spanned by coherent states lying on a circle in phase space using loop, collision, and infinity gates. In Sec. C, we relax the restriction of the coherent states lying on a circle and extend our gates to arbitrary configurations. In Sec. D, we make contact with the adiabatic theorem and Berry connections, revealing the Berry matrix for our gates. In Sec. E, we show how to implement gates of fixed photon number parity, allowing for photon loss errors to be corrected before/after the gates.

## A. Two-qudit gates

Here we extend the two-mode $\infty$ gate to arbitrary $d$. The initial adiabatic path is the same, but an extra single mode loop gate is required afterwards to make the infinity gate a true controlled phase gate.

Similar to the single-mode gate extensions stated in the main text, we let $\alpha_\nu(0) \equiv \alpha e_\nu$ be the initial coherent state values for each mode [with real non-negative $\alpha$, $e_\nu \equiv \exp(i\frac{2\pi}{d}\nu)$, and $\nu = 0, 1, \cdots, d-1$]. The two-mode jump operators from Eq. (6) of the main text now generalize to

$$F_I = (\hat{a} - \delta\alpha_{d-1})\prod_{\nu=0}^{d-2}(\hat{a} - \alpha_\nu) \quad \text{and} \quad F_{II} = (\hat{a}\hat{b} - \alpha\alpha_{d-1})(\hat{a}\hat{b} - \delta\alpha\alpha_{d-2})\prod_{\nu=0}^{d-3}(\hat{a}\hat{b} - \alpha\alpha_\nu). \quad (1)$$

When $\delta = 1$, the $d^2$ DFS basis elements $\{|\alpha_\nu\rangle\} \otimes \{|\alpha_{\nu'}\rangle\}$ are annihilated by both $F_I$ and $F_{II}$. For $\delta \neq 1$ and for sufficiently large $\alpha$, the basis elements become $\{|\alpha_\nu, X\alpha_{\nu'}\rangle\}$ and $\{|\delta\alpha_{d-1}, X\alpha_{\nu'}\rangle\}$, where $X$ follows from Tab. I. Upon adiabatic evolution of $\delta$ in an $\infty$-shaped path, the Berry phase obtained from any $|\delta\alpha_\nu\rangle$ vanishes due to each of the two loops contributing an equal area with opposite sign. The Berry phase of the non-trivial terms, which contain the component $|\delta^{-1}\alpha_\nu\rangle$, per loop $i \in \{1, 2\}$ is now

$$\psi_i = 2\alpha^2 \int_i d^2\delta|\delta|^{-4}, \quad (2)$$

which can be calculated using the corresponding Berry connections $A_{|\delta|,\arg\delta} = i\langle\delta^{-1}\alpha|\partial_{|\delta|,\arg\delta}|\delta^{-1}\alpha\rangle$ (see Appx. D). This calculation is equivalent to the geometrical reasoning provided in the main text. At the end of the gate, the phase gained from adiabatic evolution of all $d^2$ DFS states is listed in Tab. II. We can see that all DFS states of the form $|\alpha_{d-1}, \alpha_\nu\rangle$ gain the phase $\psi_1 + \psi_2$, with exception of the last state $|\alpha_{d-1}, \alpha_{d-1}\rangle$. If we now apply a loop gate on the $|\alpha_{d-1}\rangle$ state of the first oscillator with a $-(\psi_1 + \psi_2)$ phase, the last state becomes the only one to gain a phase:

$$|\alpha_{d-1}, \alpha_{d-1}\rangle + |\text{rest}\rangle \longrightarrow e^{-i(\psi_1+\psi_2)}|\alpha_{d-1}, \alpha_{d-1}\rangle + |\text{rest}\rangle, \quad (3)$$

where $|\text{rest}\rangle$ is the unaffected superposition of the remaining components $\{|\alpha_\nu, \alpha_{\nu'}\rangle\}$.

## B. Proof of universal computation

To prove universal computation of the three effective unitary gate families, we analyze their generators and show that various commutators of those generators provide a basis for the Lie algebra $\mathfrak{su}(d)$ [1]. The loop gates are generated by $d$ coherent

| mode a \ mode b | $\|X\alpha_0\rangle$ | $\|X\alpha_1\rangle$ | ... | $\|X\alpha_{d-3}\rangle$ | $\|X\alpha_{d-2}\rangle$ | $\|X\alpha_{d-1}\rangle$ |
|---|---|---|---|---|---|---|
| $\|\alpha_0\rangle$ | 1 | 1 | ... | 1 | $\delta$ | 1 |
| $\|\alpha_1\rangle$ | 1 | 1 | ... | $\delta$ | 1 | 1 |
| $\vdots$ | $\vdots$ | $\vdots$ | $\ddots$ | $\vdots$ | $\vdots$ | $\vdots$ |
| $\|\alpha_{d-3}\rangle$ | 1 | $\delta$ | ... | 1 | 1 | 1 |
| $\|\alpha_{d-2}\rangle$ | $\delta$ | 1 | ... | 1 | 1 | 1 |
| $\|\delta\alpha_{d-1}\rangle$ | $\delta^{-1}$ | $\delta^{-1}$ | ... | $\delta^{-1}$ | $\delta^{-1}$ | 1 |

Table I. Values of $X$ for each state of mode $b$ for all $d^2$ steady states of a Lindbladian with jump operators $F_I$ and $F_{II}$ from Eq. (1).

| mode a \ mode b | $\|\alpha_0\rangle$ | $\|\alpha_1\rangle$ | ... | $\|\alpha_{d-3}\rangle$ | $\|\alpha_{d-2}\rangle$ | $\|\alpha_{d-1}\rangle$ |
|---|---|---|---|---|---|---|
| $\|\alpha_0\rangle$ | 0 | 0 | ... | 0 | 0 | 0 |
| $\|\alpha_1\rangle$ | 0 | 0 | ... | 0 | 0 | 0 |
| $\vdots$ | $\vdots$ | $\vdots$ | ... | $\vdots$ | $\vdots$ | $\vdots$ |
| $\|\alpha_{d-3}\rangle$ | 0 | 0 | ... | 0 | 0 | 0 |
| $\|\alpha_{d-2}\rangle$ | 0 | 0 | ... | 0 | 0 | 0 |
| $\|\alpha_{d-1}\rangle$ | $\psi_1+\psi_2$ | $\psi_1+\psi_2$ | ... | $\psi_1+\psi_2$ | $\psi_1+\psi_2$ | 0 |

Table II. Berry phase gained for each of the $d^2$ DFS states upon completion of the $\infty$-gate.

state projections $\hat{\pi}_\nu \equiv |\alpha e_\nu\rangle\langle\alpha e_\nu|$ and can be commuted with $\hat{\chi}$ to generate $\mathfrak{su}(d)$ as follows. Superpositions of $\hat{\pi}_\nu$ provide $d-1$ linearly independent traceless diagonal elements. The linearly independent off-diagonal coherent state transitions are provided by $\frac{1}{2}d(d-1)$ elements

$$\hat{g}_{\nu\nu'} \equiv [\hat{\pi}_\nu, [\hat{\chi}, \hat{\pi}_{\nu'}]] = \frac{|\alpha e_\nu\rangle\langle\alpha e_{\nu'}|}{e^{-i\frac{2\pi}{d}(\nu-\nu')} - 1} + H.c. \qquad (4)$$

along with the $\frac{1}{2}d(d-1)$ elements $\frac{i}{2}[\hat{\pi}_\nu - \hat{\pi}_{\nu'}, \hat{g}_{\nu\nu'}]$ (both sets for $\nu \neq \nu'$). Together this makes $d^2-1$ elements, the dimension of $\mathfrak{su}(d)$. Adding in the qudit infinity gates from above then becomes sufficient for universal computation [2].

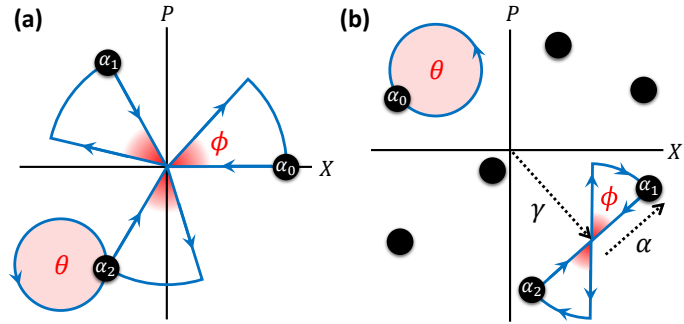

Figure 3 from main text. **(a)** Three-fold symmetric configuration of steady states $|\alpha_\nu\rangle$ of Eq. (1) with $d=3$ and depiction of a loop gate ($\theta$) acting on $|\alpha_2\rangle$ and a collision gate ($\phi$) between all states. **(b)** Arbitrary configuration of steady states for $d=7$, depicting $|\alpha_0\rangle$ undergoing a loop gate and $|\alpha_1\rangle, |\alpha_2\rangle$ (with "center of mass" $\gamma$ and relative coordinate $\alpha$) undergoing a displaced collision gate.

### C. Generalization to arbitrary $\alpha_\nu$

Here we discuss generalized versions of the single-mode symmetric 2-photon and $d$-photon gates from the main text, breaking symmetry and allowing for arbitrary values of $|\alpha_\nu\rangle$ (as long as all states remain well-separated). Fig. 3 from the main text is reproduced for convenience.

Generalized loop gates can be performed via simultaneous variation of multiple $\alpha_\nu(t)$. The procedure is exactly the same as that described in the main text for the $d = 2$ loop gate. More interestingly, collision gates away from the origin between any two (or more) coherent states can also be performed. For example, the symmetric $d = 2$ collision gate, demonstrated with the collision at the origin in Fig. 1d (see main text), can be performed anywhere in phase space. If one picks two states $|\alpha_1(0)\rangle, |\alpha_2(0)\rangle$ (see Fig. 2b) and assumes that all other states are well-separated from them, a two-state collision gate is then a displaced version of the original collision gate from Fig. 1d. To better understand this, first let $\alpha_1 = \gamma + \alpha$ and $\alpha_2 = \gamma - \alpha$ with complex relative coordinate $\alpha$ and "center of mass" $\gamma$. Then, note that the displaced cat states become (up to a global phase)

$$D_\gamma |\mu_\alpha\rangle \stackrel{|\alpha|\gtrsim 2}{\to} \frac{1}{\sqrt{2}}(|\alpha_1\rangle + (-)^\mu e^{i\operatorname{Im}\alpha_1 \alpha_2^\star}|\alpha_2\rangle),$$

where $\mu \in \{0, 1\}$ and $D_\gamma$ is the displacement operator (with $D_\gamma|0\rangle = |\gamma\rangle$). The states $D_\gamma|\mu_\alpha\rangle$ provide a basis (valid for any $\alpha_1, \alpha_2$) for the steady states of the $d = 2$ case $F = \sqrt{\kappa}(\hat{a} - \alpha_1)(\hat{a} - \alpha_2)$ from Eq. (1) of the main text, which reduces to $F = \sqrt{\kappa}(\hat{a}^2 - \alpha^2)$ when $\gamma = 0$. The collision gate thus translates into varying the amplitude and phase of $\alpha$ in the path shown in Fig. 2b while keeping $\gamma$ constant, producing the same net result as the original $d = 2$ collision gate and inducing transitions between $|\alpha_1(0)\rangle$ and $|\alpha_2(0)\rangle$. Naturally, such phase and collision gates allow for universal control for $|\alpha_\nu(0)\rangle$ at arbitrary locations in phase space (Fig. 2b), granted that $|\alpha_\nu(0)\rangle$ are well-separated from each other: $|\alpha_\nu(0) - \alpha_{\nu'}(0)| \gtrsim 4$ for all $\nu, \nu' = 0, 1, \cdots, d - 1$.

### D. Berry connections

The adiabatic theorem states that traversal of a closed loop in parameter space in the adiabatic limit (parallel transport) results in the application of a Berry matrix (more generally, a holonomy) on the initial state space, generalizing the Berry phase for non-degenerate systems. There are two adiabatic limits associated with Lindbladians: one governed by the steady states of the entire Lindbladian ([3], Sec. 6; see also [4–10]) and one governed by eigenstates of a Hamiltonian part [11]. We consider a DFS in the former limit. The Berry matrix imparted on a $d$-dimensional steady state subspace when time evolution is governed by a Lindbladian may not be equal to that imparted on the same subspace when time evolution is governed by a Hamiltonian. However, Ref. [6] showed that the Berry matrix of the Lindbladian-controlled DFS case [5] does indeed reduce to the ordinary $SU(d)$ Berry matrix for unitary systems [12]. Here we sketch another proof of this result, based on a more general derivation of the Berry matrix for generic steady state subspaces from Ref. [10]. We note that non-adiabatic HQC schemes on DFSs [13] and NSs [14] have also been proposed and that HQC (as well as general manipulations) on a DFS can alternatively be understood in terms of a generalized quantum Zeno effect [6, 15–18].

The Lindbladian Berry matrix, in general nonunitary, can be calculated via an ordered path integral of the Lindbladian Berry connections

$$\mathcal{A}^\lambda_{\mu\mu';\sigma\sigma'} = i\operatorname{Tr}\{J^\dagger_{\mu\mu'}\partial_\lambda(|\sigma_\alpha\rangle\langle\sigma'_\alpha|)\}, \tag{5}$$

where $J_{\mu\mu'}$ is the conserved quantity [19] corresponding to $|\mu_\alpha\rangle\langle\mu'_\alpha|$ and $\partial_\lambda \equiv \frac{\partial}{\partial\lambda}$. The $J_{\mu\mu'}$ determine the initial Bloch vector of the cat qudit and are known in closed form only for $d = 2$ [15]. Interestingly, $J_{\mu\mu'}$ do not participate in parallel transport for either gate and Eq. (5) reduces to its closed system counterpart – a superposition of ordinary Berry connections between cat states. This reduction holds because $J_{\mu\mu'}$ do not "cross-talk," i.e., $P_\alpha J_{\mu\mu'} P_\alpha^\perp = 0$ with $P_\alpha = \sum_{\mu=0}^{d-1}|\mu_\alpha\rangle\langle\mu_\alpha|$ the projection on the steady state subspace and $P_\alpha^\perp = 1 - P_\alpha$. This condition can be shown to imply that adiabatic parallel transport on the DFS is free from the nonunitary effects of the Lindbladian. We have shown the condition numerically for the example here; the exact proof is shown in Ref. [10]. In other words, one can verify that

$$\mathcal{A}^\lambda_{\mu\mu';\sigma\sigma'} = \delta_{\mu'\sigma'} A^\lambda_{\mu\sigma} + \delta_{\mu\sigma} A^\lambda_{\mu'\sigma'} \tag{6}$$

with $A^\lambda_{\mu\sigma} = i\langle\mu_\alpha|\partial_\lambda\sigma_\alpha\rangle$ the ordinary Berry connection [12].

The Berry connections $A^\lambda$ reduce exactly to the gate generators $\hat{\chi}, \hat{\pi}_\nu$ from the main text. For example, the symmetric collision gate in Figs. 1c and 2a arises from changes in the magnitude and phase of $\alpha$ (here complex). Therefore, $\lambda \in \{|\alpha|, \arg\alpha \equiv \varphi\}$ and a simple calculation reveals $A^{|\alpha|}_{\mu\mu'} = 0$ and

$$A^\varphi_{\mu\mu'} = -\delta_{\mu\mu'}\langle\mu_\alpha|\hat{n}|\mu_\alpha\rangle \to \begin{cases} -\delta_{\mu\mu'}\mu & |\alpha| \to 0 \\ -\delta_{\mu\mu'}|\alpha|^2 & |\alpha| \to \infty. \end{cases} \tag{7a, 7b}$$

One can see that $A^\varphi \propto \hat{\chi}$ as $|\alpha| \to 0$, confirming that $\hat{\chi}$ indeed generates the effective operation induced by the collision gate.

## E. Integration with photon loss error correction scheme

Here we discuss the parity conserving holonomic single-qubit gates; the infinity gate extends analogously. To add HQC capability to qudits protected from photon loss, we need to double the size of the DFS in order to accommodate both even and odd parity logical spaces. This is done by letting $\hat{a} - \alpha_\nu \to \hat{a}^2 - \alpha_\nu^2$ in Eq. (1) from the main text. The original $\{|\alpha_\nu\rangle\}$ basis (for sufficiently large and well-separated $\alpha_\nu$) becomes

$$|\alpha_\nu, p\rangle \sim \tfrac{1}{\sqrt{2}}(|\alpha_\nu\rangle + (-)^p |-\alpha_\nu\rangle), \tag{8}$$

where $p \in \{0, 1\}$ (mod 2) indexes the logical space and $2d$ is the dimension of the new DFS. In order to extend HQC to these states, we need to show that both gate families act identically on both logical spaces $p$. The loop gate generalizes straightforwardly. Varying $\alpha_{\nu'}(t)$ in a closed loop far away from all other $\alpha_\nu$ produces the same geometric phase for both $|\pm\alpha_{\nu'}\rangle$, so $|\alpha_{\nu'}, p\rangle \longrightarrow e^{i\theta} |\alpha_{\nu'}, p\rangle$ with $\theta$ independent of $p$. To generalize the collision gate [for which we now set $\alpha_\nu \equiv \alpha \exp(i\tfrac{\pi}{d}\nu)$], we once again need a dual basis to properly take the $\alpha \to 0$ limit. This cat state basis $\{|\mu_\alpha, p\rangle \equiv |(2\mu + p)_\alpha\rangle\}$ is obtained by letting $\mu \to 2\mu + p$ and $d \to 2d$ in Eq. (9) from the main text and plugging in Eq. (8). In the two regimes of interest,

$$|\mu_\alpha, p\rangle \sim \begin{cases} |2\mu + p\rangle & \alpha \to 0 \tag{9a} \\ \tfrac{1}{\sqrt{d}} \sum_{\nu=0}^{d-1} e^{-i\tfrac{\pi}{d}\nu(2\mu+p)} |\alpha e^{i\tfrac{\pi}{d}\nu}, p\rangle & \alpha \to \infty \, . \tag{9b} \end{cases}$$

The phase gained during the $\alpha = 0$ rotation part of the collision gate is then $\phi\langle\mu_\alpha, p|\hat{n}|\mu_\alpha, p\rangle \overset{\alpha \to 0}{\sim} \phi(2\mu + p)$. Therefore, the collision gate also acts the same way on both logical spaces, up to an overall phase of $\exp(-i\phi p)$.

---


\end{document}